\documentclass[twocolumn,graphicx,pra]{revtex4}
\usepackage{amssymb}
\usepackage{epsfig}
\usepackage{bm}
\usepackage{amsmath}
\usepackage{graphicx}
\begin{document}

\title{Time-delayed nonsequential double ionization with few-cycle laser pulses: importance of the carrier-envelope phase}
\author{C. Figueira de Morisson Faria$^1$, T. Shaaran$^{1,2}$, and M. T. Nygren$^{1,3}$}
\affiliation{$^1$Department of Physics and Astronomy, University College
London, Gower Street, London WC1E 6BT, United Kingdom\\$^2$ICFO - Institut de Ci\`{e}ncies Fot\`{o}niques,
Av. Carl Friedrich Gauss, 308860 Castelldefels (Barcelona), Spain\\$^{3}$Theoretical Physics Group, Imperial College London, South Kensington Campus, London SW7 2AZ, United Kingdom}

\date{\today}

\begin{abstract}
We perform theoretical investigations of laser-induced nonsequential
double ionization with few cycle pulses, with particular emphasis on
the dependence of the electron-momentum distributions on the
carrier-envelope phase. We focus on the recollision-excitation
with subsequent tunneling ionization (RESI) pathway, in  which a
released electron, upon return to its parent ion, gives part of its
kinetic energy to promote a second electron to an excited state. At
a subsequent time, the second electron is freed through tunneling
ionization. We show that the RESI electron-momentum distributions
vary dramatically with regard to the carrier-envelope phase. By
performing a detailed analysis of the dynamics of the two active
electrons in terms of quantum orbits, we relate the shapes and the
momentum regions populated by such distributions to the dominant set
of orbits along which rescattering of the first electron and
ionization of the second electron occurs. These orbits can be manipulated by varying the
carrier-envelope phase. This opens a wide range of possibilities for
controlling correlated attosecond electron emission by an adequate
pulse choice.
\end{abstract}
\maketitle
\section{Introduction}
Ultrashort laser pulses with durations of a few optical cycles can
reach very high intensities and, still, carry much less energy than
their longer counterparts \cite{Brabec2000}. This enables a wide
range of applications, such as controlling the collective electron
motion in condensed matter \cite{Zherebtsov2011}, manipulating
chemical reactions \cite{Schnuter}, and the generation of isolated
XUV attosecond pulses \cite{Sansone2006,Ferrari2010}. In this
pulse-length regime, the carrier-envelope phase (CEP), i.e., the
phase of the carrier frequency with respect to the pulse envelope,
dramatically influences strong-field phenomena, such as
high-harmonic generation (HHG) \cite{Brabec2000}, above-threshold
ionization (ATI) \cite{FewCycleATI} and laser-induced nonsequential
double ionization (NSDI). Specifically for NSDI, it has been shown
that the electron-momentum distributions change their shapes
considerably when this parameter varies. In fact, asymmetric
electron momentum distributions have been identified both
theoretically
\cite{NSDI2004_Faria,NSDI2004_2_Faria,Ruiz_2006,Liao_2008,Gating_2009,Micheau_2009}
and experimentally \cite{NSDIpulses_exp,Camus2012,Bergues2011} for
NSDI with few-cycle driving pulses. Most of these studies have focused
on the electron-impact ionization pathway, in
which the second electron has enough energy to overcome the second
ionization potential and both electrons reach the continuum
simultaneously.

In recent experiments, however, a similar effect has also been observed in a
parameter range for which another NSDI pathway is prevalent:
Recollision excitation with subsequent tunneling ionization (RESI)
\cite{Bergues2011,Camus2012}. RESI happens when the first electron,
upon its return to the core, promotes the second electron to an
excited bound state, from which, with laser assistance, it
subsequently tunnels \cite{B. Feuerstein}. In RESI, the first
electron leaves immediately upon rescattering, while the second
electron tunnel ionizes near the subsequent field maximum. Hence,
there is a time delay between rescattering of the first electron and
tunnel ionization of the second electron. For a comprehensive
discussion of RESI see our recent review \cite{Faria2011Rev}.

The above-mentioned experimental evidence shows asymmetric electron
momentum distributions, as functions of the momentum components
$p_{1\parallel},p_{2\parallel}$ parallel to the laser-field
polarization, whose shape varies dramatically with the CEP. For
instance, in \cite{Bergues2011} asymmetric distributions have been
observed, whose probability densities, depending on this phase, are
either stronger in the positive half axes $p_{i\parallel}=0$,
$p_{j\parallel}\geq 0$, with $i,j=1,2$ and $i\neq j$, or in the
negative half axes $p_{i\parallel}=0$, $p_{j\parallel}\leq 0$, defined in the parallel-momentum plane $p_{1\parallel}p_{2\parallel}$. As
the CEP is varied, the momentum region in which the correlated
probability density is larger shifts from one half axis to the
other. This behavior resembles to a great extent that observed for
electron-impact ionization \cite{NSDIpulses_exp}, which has been
explained by us in previous work in terms of a shift in the dominant
set of orbits along which inelastic rescattering occurs
\cite{NSDI2004_Faria,NSDI2004_2_Faria}. The main difference is that,
instead of being located either at the positive or negative momentum half axes mentioned above,
the electron-impact NSDI distributions populated either the first or the third quadrant of the $p_{1\parallel}p_{2\parallel}$ plane. Depending on the CEP range chosen, they shifted from the first to the third quadrant of the parallel-momentum plane, or vice versa.

In the present work, we address the question whether, similarly to
what happens in the direct pathway, it is possible to associate
dominant sets of trajectories to specific values of the CEP and to
specific shapes of the electron-momentum distributions. For that
purpose, we apply the analytical RESI model developed by us in
previous publications \cite{SF2010, SNF2010} to few-cycle pulses of
different CEPs. In this model, the Feynman diagram corresponding to
RESI has been considered from the outset, and the pertaining
transition amplitude has been calculated in the strong-field
approximation using the steepest descent method. A particularly
important issue in the context of RESI with few-cycle driving pulses
is that particular care must be taken with regard to causality if
the above-mentioned methods are used. This issue has been addressed
by us in a recent publication \cite{SFS2012}.

This article is organized as follows.
In Sec.~\ref{theory}, we provide a brief summary
of the method developed in \cite{SF2010, SNF2010}. In Sec.~\ref{orbitspulse}, we determine the dominant sets of orbits for the first and second electron, from the solutions of the saddle-point equations. Subsequently, in Sec.~\ref{distributions} we
compute the RESI electron-momentum distributions and analyze their CEP dependence in terms of such orbits. Finally, in Sec. \ref{Conclusions}
we state the main conclusions to be drawn from this work.
\section{Model}
\label{theory}
\subsection{Transition amplitude}
Within the strong-field approximation, the transition amplitude corresponding to RESI reads \cite{SNF2010}
\begin{eqnarray}
&&M(\mathbf{p}_{1},\mathbf{p}_{2})=\hspace*{-0.2cm}\int_{-\infty }^{\infty
}dt\int_{-\infty }^{t}dt^{^{\prime }}\int_{-\infty }^{t^{\prime
}}dt^{^{\prime \prime }}\int d^{3}k  \notag \\
&&\times V_{\mathbf{p}_{2}e}V_{\mathbf{p}_{1}e,\mathbf{k}g}V_{\mathbf{k}%
g}\exp [iS(\mathbf{p}_{1},\mathbf{p}_{2},\mathbf{k},t,t^{\prime },t^{\prime
\prime })].  \label{Mp}
\end{eqnarray}%
In Eq.~(\ref{Mp}),
\begin{eqnarray}
&&S(\mathbf{p}_{1},\mathbf{p}_{2},\mathbf{k},t,t^{\prime },t^{\prime \prime
})=  \notag \\
&&\quad E_{\mathrm{1}}^{(g)}t^{\prime \prime }+E_{\mathrm{2}}^{(g)}t^{\prime
}+E_{\mathrm{2}}^{(e)}(t-t^{\prime })-\int_{t^{\prime \prime }}^{t^{\prime }}%
\hspace{-0.1cm}\frac{[\mathbf{k}+\mathbf{A}(\tau )]^{2}}{2}d\tau  \notag \\
&&\quad -\int_{t^{\prime }}^{\infty }\hspace{-0.1cm}\frac{[\mathbf{p}_{1}+%
\mathbf{A}(\tau )]^{2}}{2}d\tau -\int_{t}^{\infty }\hspace{-0.1cm}\frac{[%
\mathbf{p}_{2}+\mathbf{A}(\tau )]^{2}}{2}d\tau  \label{singlecS}
\end{eqnarray} gives the semiclassical action while
\begin{eqnarray}
V_{\mathbf{k}g} &=&\left\langle \tilde{\mathbf{k}}(t^{\prime \prime
})\right\vert V\left\vert \psi _{1}^{(g)}\right\rangle  \notag \\
&=&\frac{1}{(2\pi )^{3/2}}\int d^{3}r_{1}V(\mathbf{r}_{1})e^{-i\tilde{\mathbf{%
k}}(t^{\prime \prime })\cdot \mathbf{r}_{1}}\psi _{1}^{(g)}(\mathbf{r}%
_{1}),  \label{Vkg}
\end{eqnarray}%
\begin{eqnarray}
V_{\mathbf{p}_{1}e,\mathbf{k}g} &=&\left\langle \tilde{\mathbf{p}}_{1}\left(
t^{\prime }\right) ,\psi _{2}^{(e)}\right\vert V_{12}\left\vert \tilde{\mathbf{\
k}}(t^{\prime }),\psi _{2}^{(g)}\right\rangle  \notag \\
&=&\frac{1}{(2\pi )^{3}}\int \int d^{3}r_{2}d^{3}r_{1}\exp [-i(\mathbf{p}%
_{1}-\mathbf{k})\cdot \mathbf{r}_{1}]  \notag \\
&&{}\times V_{12}(\mathbf{r}_{1,}\mathbf{r}_{2})[\psi _{2}^{(e)}(\mathbf{r}%
_{2})]^{\ast }\psi _{2}^{(g)}(\mathbf{r}_{2}),\quad  \label{Vp1e,kg}
\end{eqnarray}
and
\begin{eqnarray}
V_{\mathbf{p}_{2}e} &=&\left\langle \tilde{\mathbf{p}}_{2}\left( t\right)
\left\vert V_{\mathrm{ion}}\right\vert \psi _{2}^{(e)}\right\rangle  \notag
\\
&=&\frac{1}{(2\pi )^{3/2}}\int d^{3}r_{2}V_{\mathrm{ion}}(\mathbf{r}%
_{2})e^{-i\tilde{\mathbf{p}}_{2}(t)\cdot \mathbf{r}_{2}}\psi _{2}^{(e)}(%
\mathbf{r}_{2}).  \label{Vp2e}
\end{eqnarray}%
are the form factors related to the ionization of the first
electron, recollision of the first electron with excitation of the second
electron, and tunnel ionization of the second electron
[Eqs.~(\ref{Vkg}), (\ref{Vp1e,kg}) and (\ref{Vp2e}), respectively].
These form factors provide information about the binding potential
$V(\mathbf{r}_{1})$ and $V_{\mathrm{ion}}(\mathbf{r}_{2})$ ``seen" by
the first and the second electron, respectively, and about
the interaction $V_{12}(\mathbf{r}_{1},\mathbf{r}_{2})$ of
the first electron with the core.

Eq.~(\ref{Mp}) describes a process in which the first electron,
initially bound in the ground state $|\psi _{1}^{(g)}\rangle $ with
energy $E_{1}^{(g)}$, is freed by tunnel ionization at time
$t^{\prime \prime }$ into a Volkov state $|\tilde{\mathbf{k}}
(t^{\prime \prime })\rangle $. Thereafter, it propagates in the
continuum from the time $t^{\prime \prime }$ to the time $t^{\prime
}$ with intermediate momentum $\mathbf{k}$. At a time $t^{\prime }$,
it rescatters inelastically with the core and, through the
interaction $V_{12}$, excites a second electron from the ground
state $|\psi _{2}^{(g)}\rangle $ of the singly ionized target to the
state $|\psi _{2}^{(e)}\rangle$. The energies of the ground and
excited states of the singly ionized target are $E_{2}^{(g)}$ and
$E_{2}^{(e)}$, respectively. The first electron reaches the detector
with final momentum $\mathbf{p}_{1}$ immediately after rescattering.
The second electron remains bound until a later time $t$, when it is
released by tunnel ionization into a Volkov state
$|\tilde{\mathbf{p}}_{2}\left( t\right) \rangle $. It reaches the
detector with final momentum $\mathbf{p}_{2}$. In the above-stated
equations, $\tilde{\mathbf{k}}(\tau )=\mathbf{k}+\mathbf{A}(\tau )$
and $\tilde{\mathbf{p}}_{n}(\tau )=\mathbf{p}_{n}+\mathbf{A}(\tau )$
$(\tau =t,t^{\prime},t^{\prime \prime }$) in the length gauge, and
$\tilde{\mathbf{k}}(\tau )=\mathbf{k}$ and
$\tilde{\mathbf{p}}_{n}(\tau )=\mathbf{p}_{n}$ in the velocity
gauge, with $n=1,2$. In \cite{SNF2010}, we have verified that, in practice, the
results obtained in both gauges are nearly identical
 \footnote{
The length to velocity gauge transformation will introduce a shift $\mathbf{p
}\rightarrow \mathbf{p}-\mathbf{A}(t)$. This shift will effectively cancel out
with the field dressing in the Volkov states which exists in the velocity gauge, and will influence the ionization prefactors for the first and second electron. However, as tunnel ionization occurs close to the peak of the field, $A(t^{\prime\prime})\simeq 0$ in the length-gauge prefactor (\ref{Vkg}) and $A(t)\simeq 0$ in the length-gauge prefactor (\ref{Vp2e}). Hence, this will have little influence in practice.}.
\subsection{Saddle-point equations}
The transition amplitude (\ref{Mp}) is computed using the steepest descent method. Hence, one must find the values of $t,t^{\prime}, t^{\prime\prime}$ and $\mathbf{k}$ for which the action (\ref{singlecS}) is stationary, i.e., for which $
\partial _{t^{\prime \prime }}S(\mathbf{p}_{1},\mathbf{p}_{2},\mathbf{k},t,t^{\prime },t^{\prime \prime
})=\partial _{t^{\prime }}S(\mathbf{p}_{1},\mathbf{p}_{2},\mathbf{k},t,t^{\prime },t^{\prime \prime
})=\partial
_{t}S(\mathbf{p}_{1},\mathbf{p}_{2},\mathbf{k},t,t^{\prime },t^{\prime \prime
})=0,$ and $\quad \partial _{\mathbf{k}}S(\mathbf{p}_{1},\mathbf{p}_{2},\mathbf{k},t,t^{\prime },t^{\prime \prime
})=\mathbf{0}$.

The conditions upon $t^{\prime\prime}$, $\mathbf{k}$ and $t^{\prime}$ give the saddle-point equations
\begin{equation}
\left[ \mathbf{k}+\mathbf{A}(t^{\prime \prime })\right] ^{2}=-2E_{1}^{(g)},
\label{saddle1}
\end{equation}%
\begin{equation}
\mathbf{k=}-\frac{1}{t^{\prime }-t^{\prime \prime }}\int_{t^{\prime \prime
}}^{t^{\prime }}d\tau \mathbf{A}(\tau ),  \label{saddle2}
\end{equation}%
and
\begin{equation}
\left[ \mathbf{p}_{1}+\mathbf{A}(t^{\prime })^2 \right]=\left[ \mathbf{k}+\mathbf{A}%
(t^{\prime })\right] ^{2}-2(E_{2}^{(g)}-E_{2}^{(e)}), \label{saddle3}
\end{equation}
while the condition upon $t$ leads to
\begin{equation}
\lbrack \mathbf{p}_{2}+\mathbf{A}(t)]^{2}=\mathbf{-}2E_{2}^{(e)}.
\label{saddle4a}
\end{equation}
Equation (\ref{saddle1}) expresses the fact that the energy of the
first electron is conserved at the instant $t^{\prime \prime }$ at
which tunneling ionization occurs. Equation (\ref{saddle2}) fixes
the intermediate momentum of the first electron in order to
guarantee its return to its parent ion. Equation (\ref{saddle3})
states that, upon rescattering, the first electron transferred a
fraction $E_{\mathrm{exc}}=E_{2}^{(g)}-E_{2}^{(e)}$ of its kinetic energy $\left[
\mathbf{k}+\mathbf{A}(t^{\prime })\right] ^{2}/2$ upon return, and
reached the detector with final kinetic energy $\left[
\mathbf{p}_{1}+\mathbf{A}(t^{\prime })\right]^2/2$. Finally,
Eq.~(\ref{saddle4a}) gives the energy conservation for the second
electron upon tunneling, which reaches the detector with final momentum $\mathbf{p}_2$. Note that Eqs.
(\ref{saddle1}) and (\ref{saddle4a}) have no real solution, since
tunneling has no classical counterpart.
If written in terms of the electron momentum components
$(p_{n\parallel},p_{n\perp})$, $n=1,2$ parallel and perpendicular to
the laser-field polarization, Eqs. (\ref{saddle3}) and
(\ref{saddle4a}) give the kinematic constraints related to the first
and second electron, respectively. These constraints have been
discussed in detail in Refs.~\cite{SF2010,SNF2010} and \cite{SFS2012} for
monochromatic fields and few-cycle pulses, respectively, and will be
briefly mentioned here.

Explicitly, from Eq.~(\ref{saddle3}) one obtains the condition
\begin{equation}
-A(t^{\prime})-\sqrt{2\triangle E}\leq p_{1\parallel }\leq -A(t^{\prime})+\sqrt{2\triangle E},\label{regionp1}
\end{equation}
where $\triangle E$ = $E_{\mathrm{kin}}(t^{\prime },t^{\prime \prime })-%
\tilde{E}_{\mathrm{exc}}$ denotes the energy difference between the
kinetic energy $\ E_{\mathrm{kin}}(t^{\prime
},t^{\prime\prime })$ of the first electron upon return and the
energy $\tilde{E}_{\mathrm{exc}}=E_{\mathrm{exc}}+p_{1\perp}^2/2$. The latter is an
effective excitation energy, which increases with perpendicular
momentum $p_{1\perp}$. Inside the boundaries defined by Eq.~(\ref{regionp1}),
rescattering has a classical counterpart and the
probability density associated with it is significant, while,
outside those boundaries, rescattering is not classically allowed to
occur and the corresponding probability density is
vanishingly small. The largest region will be obtained for vanishing
transverse momentum $p_{1\perp}$.

This implies that (i) the region
in the parallel momentum plane determined by rescattering of the
first electron is centered around $p_{1\parallel}=-A(t^{\prime})$,
where $A(t^{\prime})$ is the vector potential at the time the first
electron returns, and (ii) the extension of this region is
determined by the energy difference $\triangle E$. In order to draw upper
bounds for such momentum regions, it is useful to assume that $p_{1\perp} = 0$ and consider the maximal kinetic energy $E^{\mathrm{max}}_{\mathrm{kin}}(t^{\prime},t^{\prime\prime})$ the first electron may have upon return. If $E^{\mathrm{max}}_{\mathrm{kin}}(t^{\prime},t^{\prime\prime})\gg E_{\mathrm{exc}}$, this region will be very large. In contrast, if
$E^{\mathrm{max}}_{\mathrm{kin}}(t^{\prime},t^{\prime\prime})\simeq E_{\mathrm{exc}}$, the energy of the returning electron will be just enough to excite the second electron. Hence, the momentum it will have subsequently to the collision will be that acquired from the field at the instant of rescattering. In terms of momentum-space constraints, this means that the region in the parallel momentum plane will collapse around $p_{1\parallel} = -A(t^{\prime})$.

For a
monochromatic field, $A(t^{\prime})=\pm 2\sqrt{U_p}$ and
$E^{\mathrm{max}}_{\mathrm{kin}}(t^{\prime },t^{\prime\prime
})=3.17U_p$ \cite{SF2010}. However, for a few-cycle pulse,
$-A(t^{\prime})$ and $E^{\mathrm{max}}_{\mathrm{kin}}(t^{\prime
},t^{\prime\prime })$ will depend on the rescattering event within
the pulse. This means that each specific rescattering event will
lead to a region in the $p_{1\parallel}p_{2\parallel}$ plane, whose
extension and center depend on the electron return time and kinetic
energy for a particular set of orbits.

Similarly, Eq.~(\ref{saddle4a}) can be rewritten as
\begin{equation}
\lbrack p_{2\parallel }+A(t)]^{2}=\mathbf{-}2E_{2}^{(e)}-p_{2\perp }^{2}.
\label{saddle4perp}
\end{equation}
The above-stated expression shows that there will be a large drop in
the yield with increasing transverse momentum, as
$\mathbf{p}_{2\perp }$ effectively widens the potential barrier
through which the second electron must tunnel. The
electron tunnels most probably at the laser field maxima, for which
$A(t)\ll 1$ and $p_{2\parallel}\simeq 0$. Hence, the momentum-space
conditions for which ionization of the second electron is most
probable lie approximately at $(p_{2\parallel },p_{2\perp })=(0,0)$.
For a few-cycle pulse, depending on the specific tunnel ionization
event, the barrier will be narrower or wider.

 In summary, the constraints upon the momentum of the second electron
 determine the position of the cross-shaped distributions at the axes $p_{n\parallel}=0$ in the
 parallel-momentum plane, while those related to the momentum of
 the first electron determine the regions along these axes that will be populated, i.e., whether they will be long or short. Clearly,
 for a few-cycle pulse this length will vary with each particular
 event. 
 
\section{Quantum-orbit analysis}
\label{orbitspulse}

In  this section, we will perform a quantum-orbit analysis of the
problem, with emphasis on how the dominant sets of orbits change
with the CEP. The concept of ``quantum orbits" is
based on the fact that the solutions of the saddle-point equations
can be related to the classical orbits of an electron in a field
and, still, retain information on quantum aspects such as tunneling
and interference (for a broad overview see \cite{Science2001}).

We employ the linearly polarized few-cycle pulse
$\mathbf{E}(t)=-\partial _{t}\mathbf{A}(t)/dt$,
where the vector potential $\mathbf{A}(t)$ is given by
\begin{equation}
\mathbf{A}(t)=2\sqrt{U_{p}}\sin ^{2}\left( \frac{\omega
t}{2N}\right) \sin (\phi +\omega t)\hat{e}_{z}.  \label{pulse}
\end{equation}%
In Eq.~(\ref{pulse}), $N$ denotes the number of cycles in the pulse, $\omega$ is the field frequency,
$\phi $ the carrier-envelope phase, $U_{p}=E^2_0/(4\omega^2)$ is the
ponderomotive energy, $E_0$ is the field amplitude and $\hat{e}_{z}$ the polarization vector.
Throughout, we choose the number of cycles as $N=4.3$ and
$\phi=\phi_1-\phi_0$, where $\phi_0=60^{\circ}$ is an offset value.
This is well within the parameter range employed in experiments
~\cite{Bergues2011}. Throughout, we will refer to the phase $\phi_1$
without the offset value to facilitate a comparison with the
existing literature. The bound-state energies taken correspond
to argon. Initially, both electrons are bound in the $3p$ state. The
first electron recollides with the core, exciting a second electron
to the $4s$ state.

\subsection{Approximate ionization and rescattering times}

\begin{figure}[tbp]
\hspace*{-0.5cm}\includegraphics[width=9.5cm]{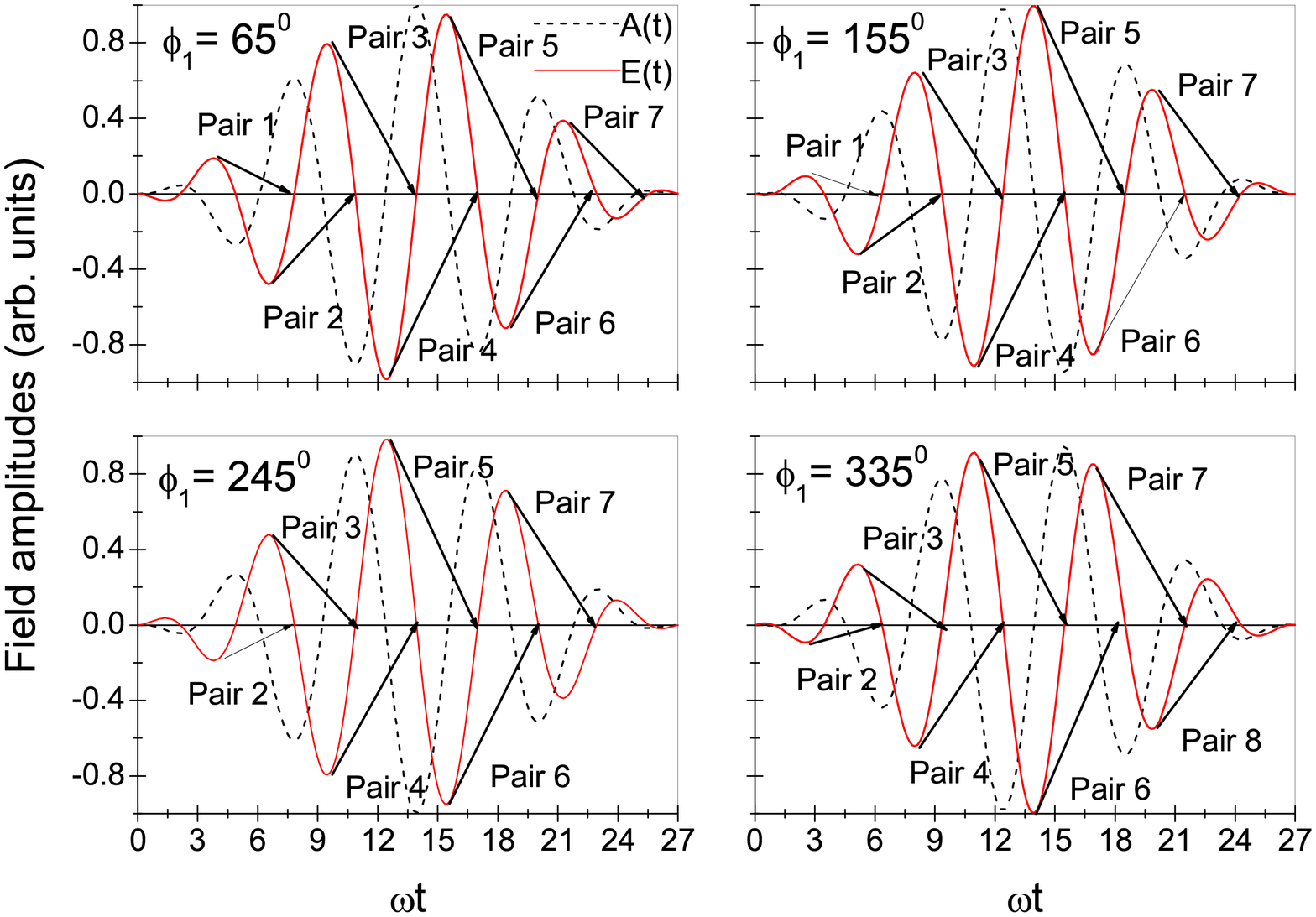}
\caption{Schematic representation of the electric field
$\mathbf{E}(t)$ and the corresponding vector potential
$\mathbf{A}(t)$, for a short pulse of $N=4.3$ cycles, whose shape is
defined by Eq.~(\ref{pulse}) and whose carrier-envelope phases are
the same as in the previous figure. The arrows indicate the
approximate classical times around which the \textit{first} electron
leaves, in case it returns at a field crossing. The pairs of orbits
are indicated by the labels Pair $n$, where $n$ ranges from 1 to 8.
The fields have been normalized to $E(t)/E_{0}$ and $A(t)/A_{0}$,
where $E_{0}$, $A_{0}$ denote the field amplitudes. The carrier
envelope phases are
$\protect\phi_1=65^{\circ}(\protect\phi=5^{\circ})$,
$\protect\phi_1=155^{\circ} (\protect\phi=95^{\circ})$,
$\protect\phi_1=245^{\circ} (\protect\phi=185^{\circ})$ and
$\protect\phi_1=335^{\circ} (\protect\phi=275^{\circ}$).}
\label{Fig2}
\end{figure}

We will now identify the relevant sets of orbits for the pulse
(\ref{pulse}). In Fig.~\ref{Fig2}, we indicate the approximate ionization
and rescattering times for the first electron, for the values of the
CEP employed in this article. These sets of times
are associated with the real part of the complex times
$t^{\prime\prime}$ and $t^{\prime}$, obtained from the solutions of
the saddle-point equations (\ref{saddle1})-(\ref{saddle3}).  Such
solutions always occur in pairs, which, physically, correspond to
the fact that the first electron may return along a shorter and a
longer orbit. In the figure, one may identify up to eight pairs of
orbits, whose ionization and rescattering times vary with the
CEP. In the following, we will refer to these
pairs as Pair $n(e_1)$, with $n=1,...,8$. For each pair, the
electron will leave most probably at a local maximum and return most probably at
the subsequent crossing.  The most relevant pairs are those near the
center of the pulse, as the field intensity in this case is higher.

\begin{figure}[tbp]
\hspace*{-0.5cm}\includegraphics[width=9.5cm]{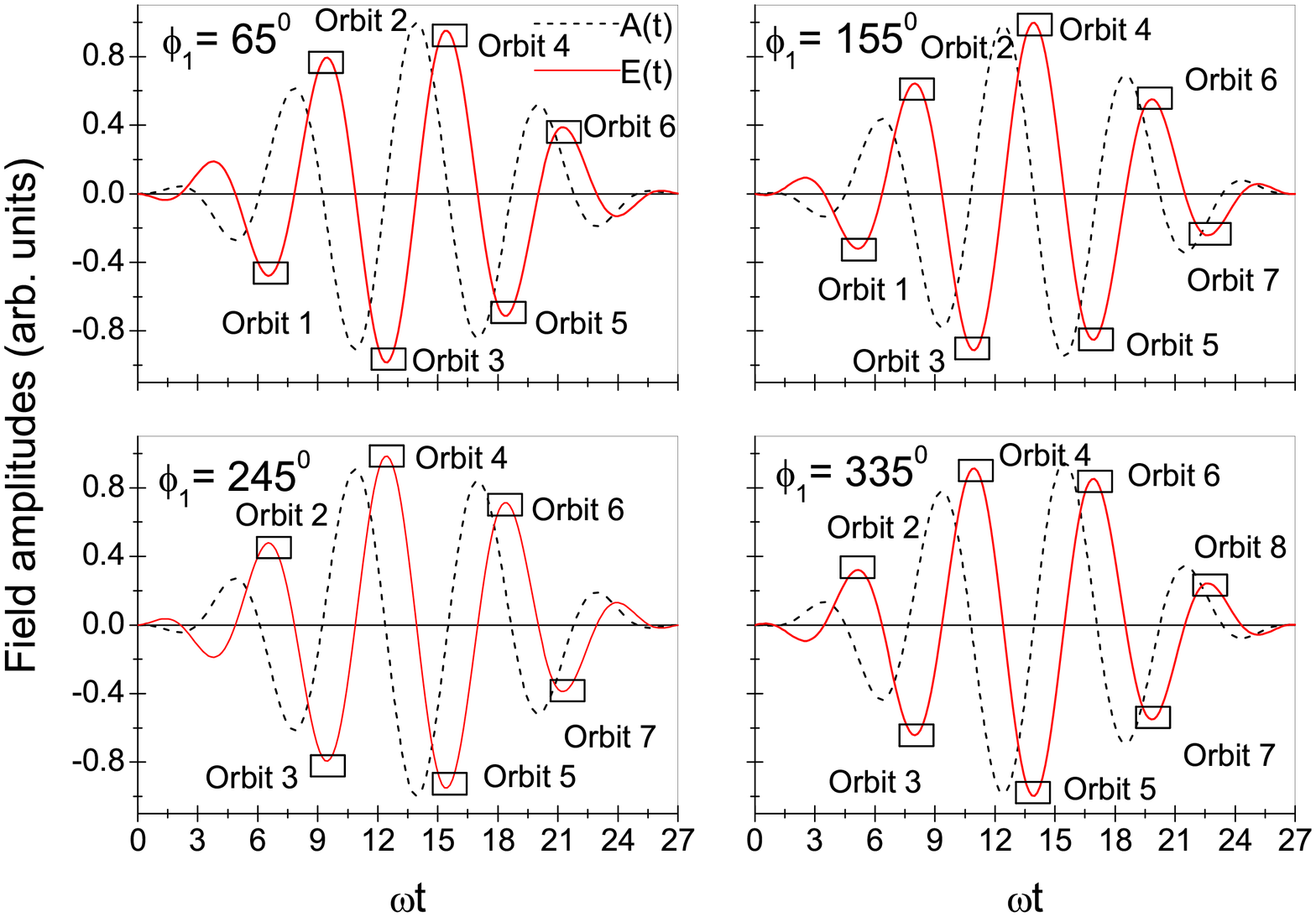}
\caption{Schematic representation of the same electric field $\mathbf{E}(t)$ and
the corresponding vector potential $\mathbf{A}(t)$ as in the previous figures, but highlighting the times around which ionization of the \textit{second} electron is expected to occur, indicated by the squares in the figure. The orbits are indicated by the labels Orbit $n$, where $n$ ranges from 1 to 8. The fields have been normalized to $E(t)/E_{0}$
and $A(t)/A_{0}$, where $E_{0}$, $A_{0}$ denote the field amplitudes.}
\label{Fig3}
\end{figure}

Fig.~\ref{Fig3} shows the approximate times at which the second
electron tunnels. These times are located around the field maxima,
and also strongly depend on the CEP. The
corresponding orbits will be referred to as Orbit $n(e_2)$, with $n$
ranging from 1 to 8. Once more, the orbits for which tunneling is
expected to be most prominent are close to the center of the pulse.
Note that the second electron cannot tunnel from an excited state
before the first electron
rescatters. This implies that, for instance,
ionization related to Orbit $2(e_2)$ can only be caused by Pair
$1(e_1)$, or ionization related to Orbit $3(e_2)$ by Pairs $1(e_1)$
and $2(e_1)$. Even though this sounds obvious, a rigorous treatment
of causality can be a non-trivial issue, especially within the
context of the steepest descent method, and requires an extensive
modification of the contours to be taken into account. For a
detailed discussion, see our previous work \cite{SFS2012}.
\subsection{Solutions of the saddle-point equations}
\subsubsection{First electron}
We will now bring together the intuitive picture discussed above and the solutions
of the saddle-point equations. We will start by focusing on the
first electron, for which the saddle-point equations
(\ref{saddle1})--(\ref{saddle3}) give the complex ionization and
rescattering times. We will study Pairs $3(e_1)$, $4(e_1)$ and
$5(e_1)$, as, for a wide range of CEPs, such pairs are expected to lead to the most relevant contributions for
the specific pulse chosen. For the cases studied in this section, the saddle-point solutions have
been obtained for vanishing momenta $p_{1\perp}$ and different CEP
values. 
\begin{figure}[tbp]
\hspace*{-0.5cm}\includegraphics[width=9.5cm]{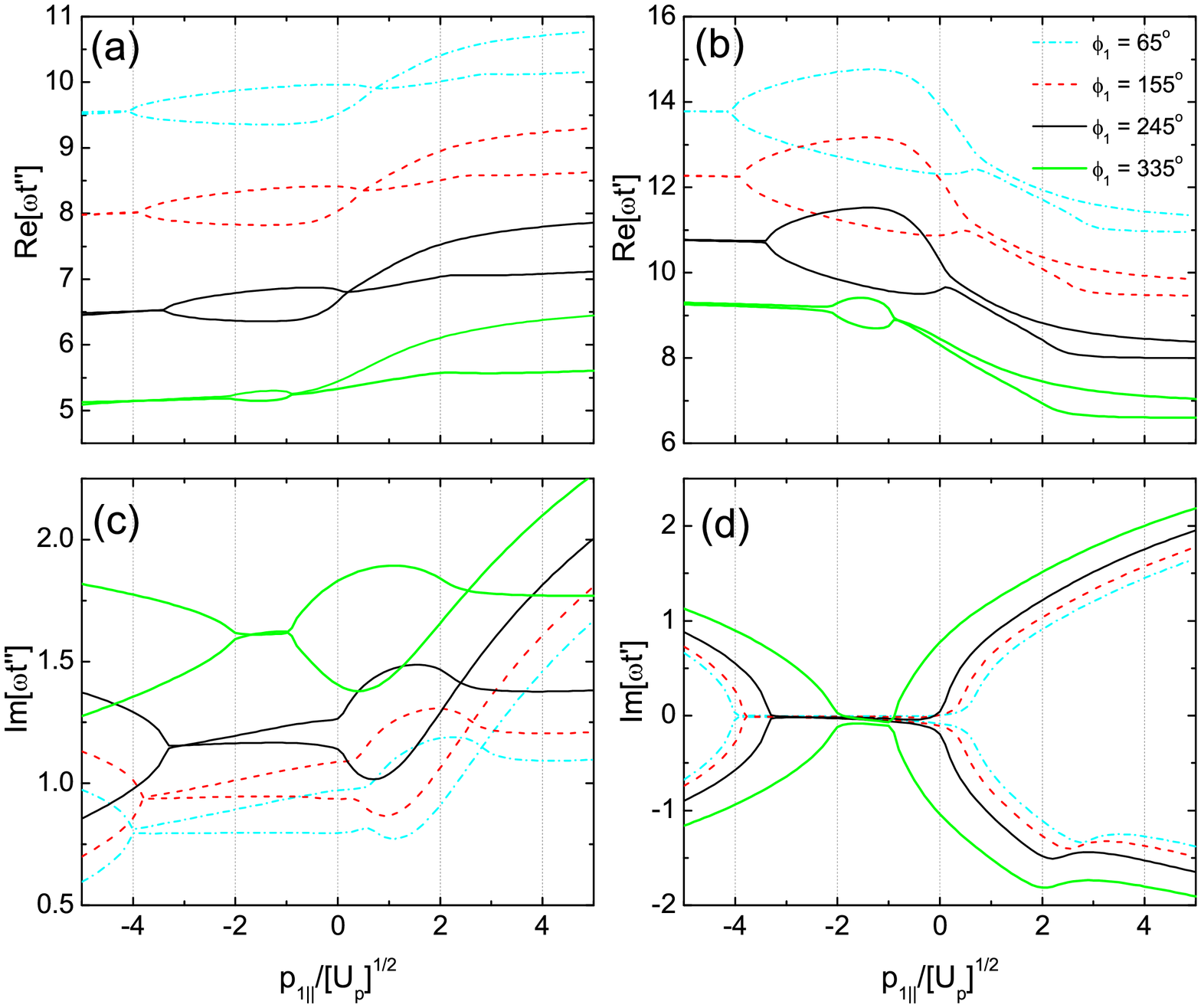}
\caption{Real and imaginary parts of the complex ionization and rescattering times $t^{\prime\prime}$ and $t^{\prime}$ of the first electron obtained from the solutions of the saddle-point Equations (\ref{saddle1})--(\ref{saddle3}) corresponding to Pair 3 in \ref{Fig2}, for vanishing transverse momentum $p_{1\perp}=0$. Panels (a) and (b) give the real parts of such solutions, and panels (c) and (d) the imaginary parts. The bound-state energies correspond to argon assuming a $3p\rightarrow 4s$ excitation ($%
E_{1g}=0.58$ a.u., $E_{2g}=1.02$ a.u. and $E_{2e}=0.40 $ a.u.) in a
few-cycle pulse of frequency $\protect\omega=0.06085$ a.u. and
intensity $I=3 \times 10^{14}\mathrm{W}/\mathrm{cm}^2$.}
\label{Fig4}
\end{figure}

Fig.~\ref{Fig4} displays the real and imaginary parts of the
above-mentioned solutions, for Pair $3(e_1)$. As an overall feature,
the real parts of the ionization and return times are centered
around $p_{1\parallel}=-A(t')$, which is in the negative parallel
momentum region (see Figs~\ref{Fig4}(a) and (b), respectively). This
is the most probable momentum for the first electron upon
rescattering.
Furthermore, $\mathrm{Re}[\omega t^{\prime\prime}]$ and
$\mathrm{Re}[\omega t^{\prime}]$ almost coalesce at two specific
values of the parallel momentum components $p_{1\parallel}$. These
are the minimum and the maximum momentum values for which
rescattering exhibits a classical counterpart. Outside this region,
this process is forbidden and the corresponding transition amplitude
is exponentially decaying.

The figure shows that the ionization and rescattering times, the
center and the extension of the classically allowed region strongly
depend on the CEP. In particular, there is a very
good agreement between the center of the region shown in
Figs.~\ref{Fig4}(a) and (b) and the ionization and return times
indicated by arrows in Fig.~\ref{Fig2}, which were identified
employing classical arguments. As the CEP
increases, the real parts of such times move towards the pulse
turn-on. This will lead to a decrease in the extension of the
classically allowed region from $-4 \sqrt{U_p}\leq
p_{1\parallel}\leq 0.7 \sqrt{U_p}$ for $\phi_1=65^{\circ}$ to
$-2\sqrt{U_p}\leq p_{1\parallel}\leq -0.8 \sqrt{U_p}$ for
$\phi_1=335^{\circ}$. This can be physically understood by direct
inspection of Fig.~\ref{Fig2}, and by bearing in mind that the
classical limit of the kinetic energy of the first electron upon return is given by
$E^{\mathrm{(cl)}}_{\mathrm{kin}}=(A(t^{\prime})-A(t^{\prime\prime}))^2/2$\footnote{Note
that, if the bound-state energy $E^{(g)}_{1}\rightarrow 0$, i.e., in
the classical limit, the saddle-point equation (\ref{saddle1}) gives
$\mathbf{k}=-\mathbf{A}(t^{\prime\prime})$.}. For
$\phi_1=65^{\circ}$, at the ionization time related to Pair
$3(e_1)$, $A(t^{\prime\prime})\simeq0$, and, at the rescattering
time, the vector potential $A(t^{\prime})$ is around its peak value
$A_0$.  Hence, the kinetic energy of the first electron upon return is high. This implies that, for this phase, the contributions of Pair $3(e_1)$ will populate a large momentum region. As the CEP is increased up to
$\phi_1=335^{\circ}$, the instantaneous vector potential
$A(t^{\prime})$ decreases to less than $0.8 A_0$. Furthermore, due
to the lack of monochromaticity of the field, $A(t^{\prime\prime})$
is no longer vanishing near the pulse turn-on. Both effects lead to a decrease in this momentum region.
 Apart from that, the center of the classically allowed region, which is determined by the most
 probable momentum the electron may have upon return, moves from roughly $-2\sqrt{U_p}$ to $-1.4\sqrt{U_p}$.
This is due to the fact that, close to the peak of the pulse, the former value is a good approximation
for $-A(t^{\prime})$, while, near the edges of the pulse, its lack of monochromaticity plays an
increasingly important role.

The imaginary parts of $t^{\prime\prime}$ and $t^{\prime}$, shown in
the remaining panels of Fig.~\ref{Fig4}, give valuable information
on whether the process in question is allowed or forbidden, or on
the overall probability related to a specific process, e.g.,
tunneling ionization. For instance, $\mathrm{Im}[\omega
t^{\prime\prime}]$, displayed in Fig.~\ref{Fig4}(d), sheds some
light on how the width of the potential barrier that the electron must
overcome in order to reach the continuum varies for Pair $3(e_1)$,
with regard to the CEP. The larger
$\mathrm{Im}[\omega t^{\prime\prime}]$ is, the wider is the
potential barrier through which the electron must tunnel
\footnote{Roughly speaking, the tunneling amplitude is proportional
to $\exp{-\mathrm{Im}[t^{\prime\prime}]}$. If, for instance,
$\mathrm{Im}[t^{\prime\prime}]$ increases by a factor of two, this
means that there will be a decrease in two orders of magnitude in
this amplitude.}. The picture shows a marked increase in
$\mathrm{Im}[t^{\prime\prime}]$ as the CEP varies
from $\phi_1=65^{\circ}$ to $\phi_1=335^{\circ}$. This can be understood with the help of Fig.~\ref{Fig2}. According to this figure, for $\phi_1=65^{\circ}$, the first
electron tunnels near a local maximum for which the instantaneous
electric field is $0.8 E_0$. As
the CEP increases, the local maximum associated
with tunnel ionization for Pair $3(e_1)$ decreases down to less than
$0.4 E_0$ for $\phi_1=335^{\circ}$. Hence, the potential barrier widens and the
contributions of this specific pair become less and less relevant.
Note that $\mathrm{Im}[\omega t^{\prime\prime}]\neq0$ throughout, as
tunneling is not classically allowed.

Finally, Fig.~\ref{Fig4}(d) shows the behavior of
$\mathrm{Im}[\omega t^{\prime}]$, which is associated to the
rescattering time, with regard to the CEP. In
contrast to what happens to $\mathrm{Im}[\omega t^{\prime\prime}]$,
this imaginary part
 vanishes at the momentum range between the values of $p_{1\parallel}$ for which
 $\mathrm{Re}[\omega t^{\prime\prime}]$ and $\mathrm{Re}[\omega t^{\prime}]$ almost
 coalesce. This is related to the fact that, in this region, rescattering is classically
 allowed. This region decreases in extension for increasing values of $\phi_1$. Once more,
this reflects the fact that Pair $3(e_1)$ loses relevance.

\begin{figure}[tbp]
\hspace*{-0.5cm}\includegraphics[width=9.5cm]{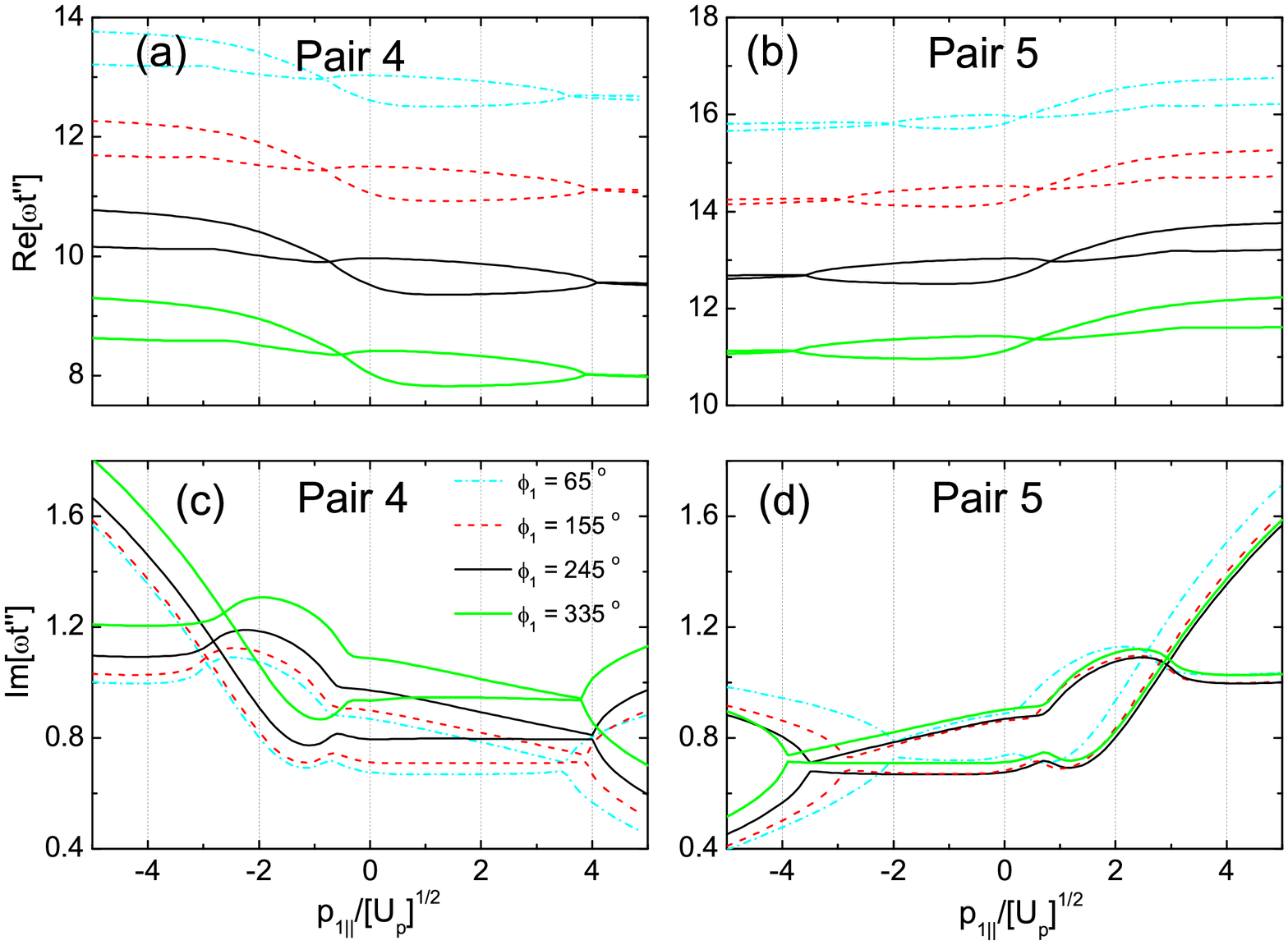}
\caption{Real and imaginary parts of the complex ionization times
$t^{\prime\prime}$ of the first electron obtained from the solutions
of the saddle-point Equations (\ref{saddle1})--(\ref{saddle3})
corresponding to Pairs 4 and 5 in \ref{Fig2}, for vanishing
transverse momentum $p_{1\perp}=0$. Panels (a) and (b) give the real
parts of such solutions, and panels (c) and (d) the imaginary parts.
The field and atomic parameters are the same as in the previous
figures.} \label{Fig5}
\end{figure}
The above-mentioned loss of relevance does not happen to all pairs
of orbits, but will depend strongly on how the corresponding start
and return times are located within the pulse. In fact, it may occur
that, as the CEP increases, a specific set of
trajectories becomes dominant, or even remains relatively stable. In
order to understand this issue, it suffices to analyze the real and
imaginary parts of the ionization time $t^{\prime\prime}$. An
example is provided in Fig.~\ref{Fig5} for Pair $4(e_1)$ and Pair
$5(e_1)$. The contributions of the former pair to the NSDI
distributions remain relatively stable, while the latter pair
increases in relevance within the CEP range studied. The real and
imaginary parts of the complex ionization time $t^{\prime\prime}$
are depicted in Fig.~\ref{Fig5} for both orbits.

Fig.~\ref{Fig5}(a) shows that, for Pair $4(e_1)$, the classically
allowed region, centered at a positive parallel momentum, remains
roughly the same throughout. This happens because the absolute value
of the vector potential $A(t^{\prime})$, and hence the kinetic energy obtained by the electron upon
return, remains quite large for the CEP range studied. The behavior
of Pair $5(e_1)$, however, displayed in Fig.~\ref{Fig5}(b), is
markedly different. For this pair, the extension of the classically
allowed region increases substantially as the CEP
increases. This is expected as, in this case, the approximate
ionization and rescattering times in Fig.~\ref{Fig2} move from the turn off
to the center of the pulse with increasing CEP. Consequently, the vector
potential $A(t^{\prime})$ at the instant of rescattering
increases from approximately $0.4A_0$ to $A_0$. This leads to a
substantial increase in the electron kinetic energy upon return.

 Another important issue determining the dominance of a pair of orbits is
 the width of the potential barrier through which the first electron tunnels,
 which can be roughly inferred from $\mathrm{Im}[\omega t^{\prime\prime}]$.
 For Pair $4(e_1)$, $\mathrm{Im}[\omega t^{\prime\prime}]$ remains relatively stable throughout,
 as shown in Fig.~\ref{Fig5}(c). An inspection of Fig.~\ref{Fig2} shows that,
 indeed, the instantaneous electric field strength $|E(t^{\prime\prime})|$
 related to Pair $4(e_1)$ is rather large. Hence, one expects the corresponding
 potential barrier to be quite narrow. For increasing CEP,
 $\mathrm{Im}[t^{\prime\prime}]$ increases slightly. This is due to the fact
 that $\mathrm{Re}[\omega t^{\prime\prime}]$ related to this pair
 moves towards the pulse turn on and the potential barrier at the ionization time $t^{\prime\prime}$
 becomes slightly wider. A similar analysis can be performed in the imaginary parts of the
saddle-point solutions for Pair $5(e_1)$. Fig.~\ref{Fig5}(d) shows
that such imaginary parts are comparable to those observed for Pair
$4(e_1)$ for a wide range of phases. Hence, we expect both sets of orbits to
compete, until Pair $5(e_1)$ becomes dominant.

\subsubsection{Second electron}

 \begin{figure}[tbp]
\hspace*{-1.5cm}\includegraphics[width=11cm]{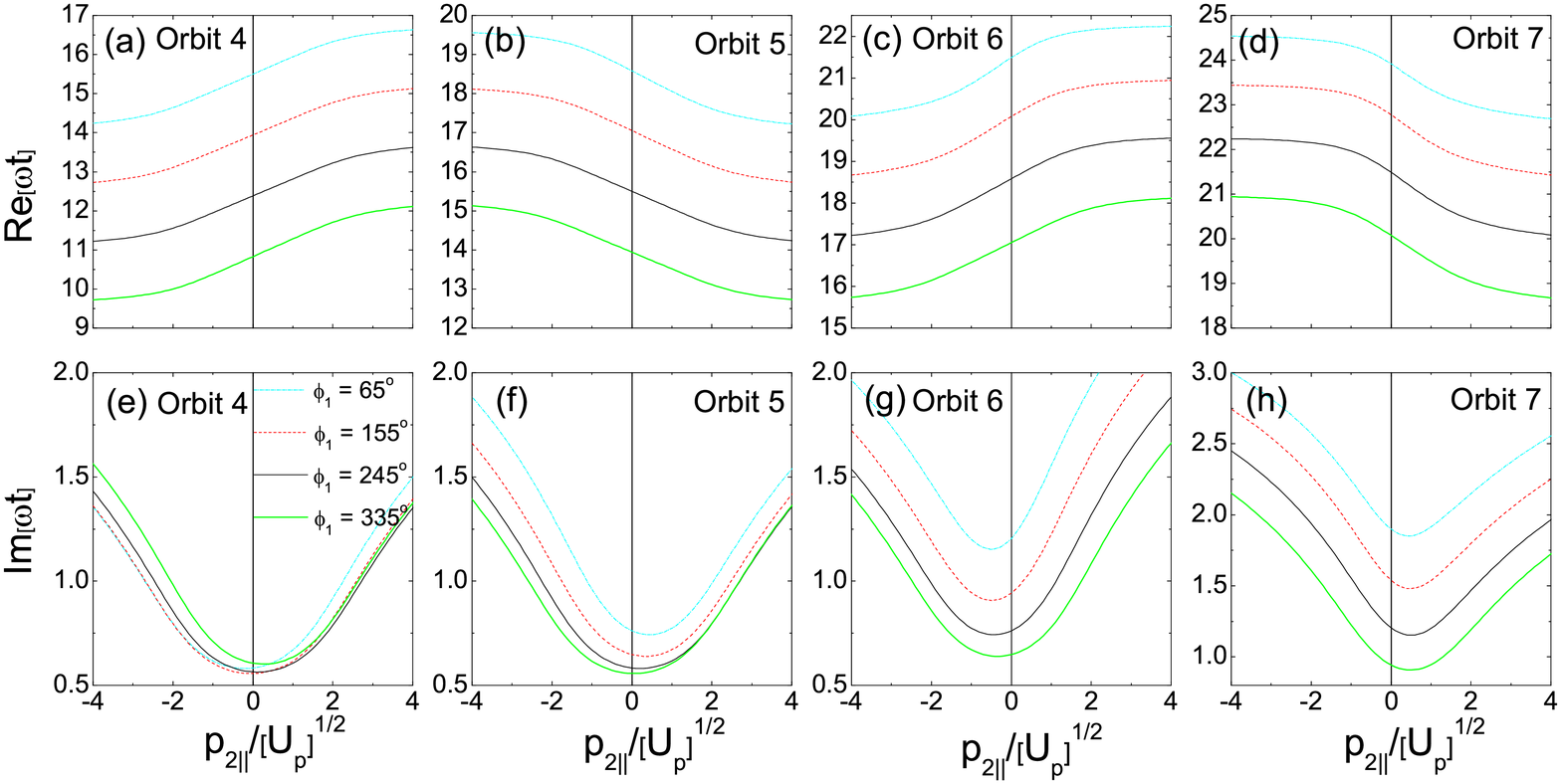} \caption{Real
and imaginary parts of the complex ionization  times $t$ of the
second electron obtained from the solutions of the saddle-point
Equation (\ref{saddle4a})) corresponding to Orbits 4 to 7 in
\ref{Fig3}, for vanishing transverse momentum $p_{2\perp}=0$. Panels
(a) to (d) give the real parts of such solutions, while panels (e)
to (h) depict the imaginary parts.} \label{Fig6}
\end{figure}
 For the second electron, a set of orbits will only be dominant
if it is excited in such a way that its subsequent
tunneling probability is high.  In order to obtain this information, we will look at the tunnel
ionization times of the second electron, which can be obtained by solving
the saddle-point equation (\ref{saddle4a}). The real and imaginary
parts of these solutions are displayed in Fig.~\ref{Fig6} (upper and
lower panels, respectively), for vanishing transverse momentum
$p_{2\perp}=0$. We only consider Orbits $n(e_2)$ for $n$ ranging
from $4$ to $7$. The preceding orbits, i.e., Orbits $n(e_2)$ with
$n=1,2,3$, will not play an important role, as they violate
causality with regard to Pair $3(e_1)$. Note that, for the second
electron, the orbits are well separated for all momentum ranges.
There is also no classically allowed region, as tunneling is an
intrinsically quantum mechanical phenomenon. Hence,
$\mathrm{Im}[\omega t]\neq 0$ throughout. Once more, one may draw an
intuitive picture by relating the solutions in Fig.~\ref{Fig6} to
the simplified arguments illustrated in Fig.~\ref{Fig3}.

For all phases, we observe that, for parallel momentum
$p_{2\parallel}\simeq0$, $\mathrm{Re}[\omega t]$ corresponds to the
times for which the pulse has local extrema, marked by the
rectangles in Fig.~\ref{Fig3}. These extrema vary from the center
until the pulse turn off, and correspond to the most probable
ionization times for the second electron, i.e., when the potential
barrier is narrowest for a specific orbit. A maximal ionization
probability near $p_{2\parallel}=0$ and its relation to the
narrowest potential barrier can also be inferred by inspecting
$\mathrm{Im}[\omega t]$, which exhibits a minimum around this value
(see lower panels in the figure). The above-stated features, i.e., the electron tunneling most
 probably at peak fields and a corresponding minimum at $\mathrm{Im}[t]$
 at such times, are also observed for a monochromatic field
 \cite{SF2010}.

However, due to the lack of
monochromaticity in the pulse, there are sometimes small deviations
from $p_{2\parallel}=0$, especially if the orbits are near the pulse turn on and off. This feature can be seen very clearly, for instance, in Fig. ~\ref{Fig6}(h), which corresponds to a tunnel ionization event close to the pulse turn off, i.e., to Orbit $7(e_2)$. Furthermore, $\mathrm{Im}[\omega t]$ is not symmetric with regard to its value at $p_{2\parallel}=0$. This would be the case for a monochromatic field, as all cycles, and hence the regions around each local field maximum, would be identical for a continuous wave \cite{SF2010}. In a few-cycle pulse, however, one expects the potential barrier through which the second electron must tunnel to narrow, or to widen, as the ionization times approach or distance themselves from the center of the pulse, respectively. This leads to asymmetries in $\mathrm{Im}[\omega t]$ for positive and negative momenta. In the specific pulse discussed in this  work, for Orbits $5(e_2)$ and $7(e_2)$ [Figs.~\ref{Fig6}(f) and (h), respectively], $\mathrm{Im}[\omega t]$ decreases for $p_{2\parallel}>0$, in comparison to the region for which $p_{2\parallel}<0$, while, for Orbit $6(e_2)$, the opposite behavior holds.  This asymmetry can be understood by inspecting the respective upper panels, together with Fig.~\ref{Fig3}. For Orbits $5(e_2)$ and $7(e_2)$, if the electron leaves before the corresponding field maxima, its momentum will be positive (see Figs.~\ref{Fig6}(b) and (d), respectively). An inspection of Fig.~\ref{Fig3} shows that these temporal regions are located closer to the center of the pulse, compared to the time regions subsequent to the local field maxima. In contrast, for Orbit $6(e_2)$, earlier ionization times, and a larger proximity to the central region of the pulse, correspond to negative momenta. For orbit $4(e_2)$, the above-mentioned asymmetry varies, as shown in Fig.~\ref{Fig6}(e). This happens because, throughout, the ionization times are very close to the pulse center. Hence, whether ionization times prior or subsequent to the local maxima will correspond to higher or lower local field intensities will depend on the carrier-envelope phase.

 The dominant sets of orbits in a few cycle pulse can be identified by analyzing Fig.~\ref{Fig6}.
 As a general feature, as $\phi_1$ increases, the real parts
 of the ionization times move towards lower values, as shown
 in Figs.~\ref{Fig6}.(a)--(d). This is consistent with Fig.~\ref{Fig3}
 and with the previous analysis, performed for the first electron. Some
 orbits, such as Orbits $5(e_2)$ -- $7(e_2)$, move towards
 the center of the pulse, while others, such as Orbit $4(e_2)$,
 move away from it.  This behavior, and its consequences for the shape of the
 electron-momentum distributions, can be inferred by
 analyzing the imaginary parts of the ionization times, displayed in
Figs.~\ref{Fig6}.(e)--(h). These panels show that the most
 important orbits for the second electron will be Orbits $4 (e_2)$
 and $5(e_2)$. This is due to the fact that, for these orbits,
 near $p_{2\parallel}=0$, $0.5\leq\mathrm{Im}[\omega t]\leq 0.75$,
 while, for the remaining orbits, $\mathrm{Im}[\omega t]$ is in general
 larger. This implies that the potential-energy barriers through which
 the electron must tunnel are narrower for Orbits $4(e_2)$ and $5(e_2)$.
 Interestingly, $\mathrm{Im}[\omega t]$ remains stable for Orbit $4(e_2)$
 throughout. This may be understood by inspecting the local maximum of $E(t)$
 in Fig.~\ref{Fig3} related to this orbit. The figure shows that
 the instantaneous field strength at this maximum remains located near $0.8E_0$
 for the CEP range considered.
 In contrast, for Orbit $5(e_2)$, $\mathrm{Im}[\omega t]$
 decreases systematically from around $0.75$ to $0.5$ as the phase
 varies from from $\phi_1=65^{\circ}$ to $\phi_1=335^{\circ}$
 (see Fig.~\ref{Fig6}(f)). This occurs because the
 instantaneous field strength $|E(t)|$ related to this orbit increases from
 less than $0.5E_0$ to almost the full amplitude $E_0$ in this phase
interval, as shown in Fig.~\ref{Fig3}.

For the remaining pairs of
orbits, there is always a decrease in $\mathrm{Im}[\omega t]$ with
regard to increasing CEP values. This is due to the fact that
the real parts of such times move from the pulse turn off towards
the center of the pulse, so that the effective potential barrier
becomes narrower. In particular, we expect vanishingly small
contributions throughout for Orbit $7(e_2)$, as in this case
$\mathrm{Im}[\omega t]$ is much larger than for the other orbits.

\section{Electron-momentum distributions}
\label{distributions}
In this section, we will apply the information gained from the quantum-orbit analysis in order to determine the regions in momentum space that should be occupied by the electron-momentum distributions. These predictions will be compared with the outcome of the actual computations, performed by setting $V_{\mathbf{p}_1e, \mathbf{k}g}=const.$, $V_{\mathbf{p}_2e}=const.$ This guarantees that both the excitation and ionization prefactors do not introduce any momentum bias, and the distributions represent the momentum constraints discussed in the previous section.
\subsection{Partial distributions}
For simplicity, we will commence by analyzing the partial momentum distributions
\begin{equation}
F^{(1)}(p_{1\parallel})=\int{|M^{(1)}(\mathbf{p}_{1})|^2}d^2p_{1\perp}\label{Mp1par}
\end{equation}
and
\begin{equation}
F^{(2)}(p_{2\parallel})=\int{|M^{(2)}(\mathbf{p}_{2})|^2}d^2p_{1\perp},\label{Mp2par}
\end{equation}
where the partial transition amplitudes for the first and the second electron read
\begin{equation}
M^{(1)}(\mathbf{p}_{1})=\int_{-\infty }^{\infty}\hspace*{-0.3cm}dt''
\hspace*{-0.1cm}\int_{t''}^\infty\hspace*{-0.3cm}dt'\hspace*{-0.1cm}
\int d^{3}kV_{\mathbf{p}_{1},\mathbf{k}}^{(eg)}V_{
\mathbf{k}}^{(g)}e^{iS_1(\mathbf{p}_{1},\mathbf{k},t'',t')}  \label{Mp1}
\end{equation} and
\begin{equation}
M^{(2)}(\mathbf{p}_{2})=\int_{-\infty}^{\infty }\hspace*{-0.2cm}dtV_{
\mathbf{p}_{2}}^{(e)}e^{iS_2(\mathbf{p}_{2},t)},  \label{Mp2}
\end{equation}
respectively and the transverse momentum components are integrated over. Note that, due to causality, the total ionization probabilities are not the product of such functions \cite{SFS2012}.

\begin{figure}
\hspace*{-1.5cm}\includegraphics[width=12cm]{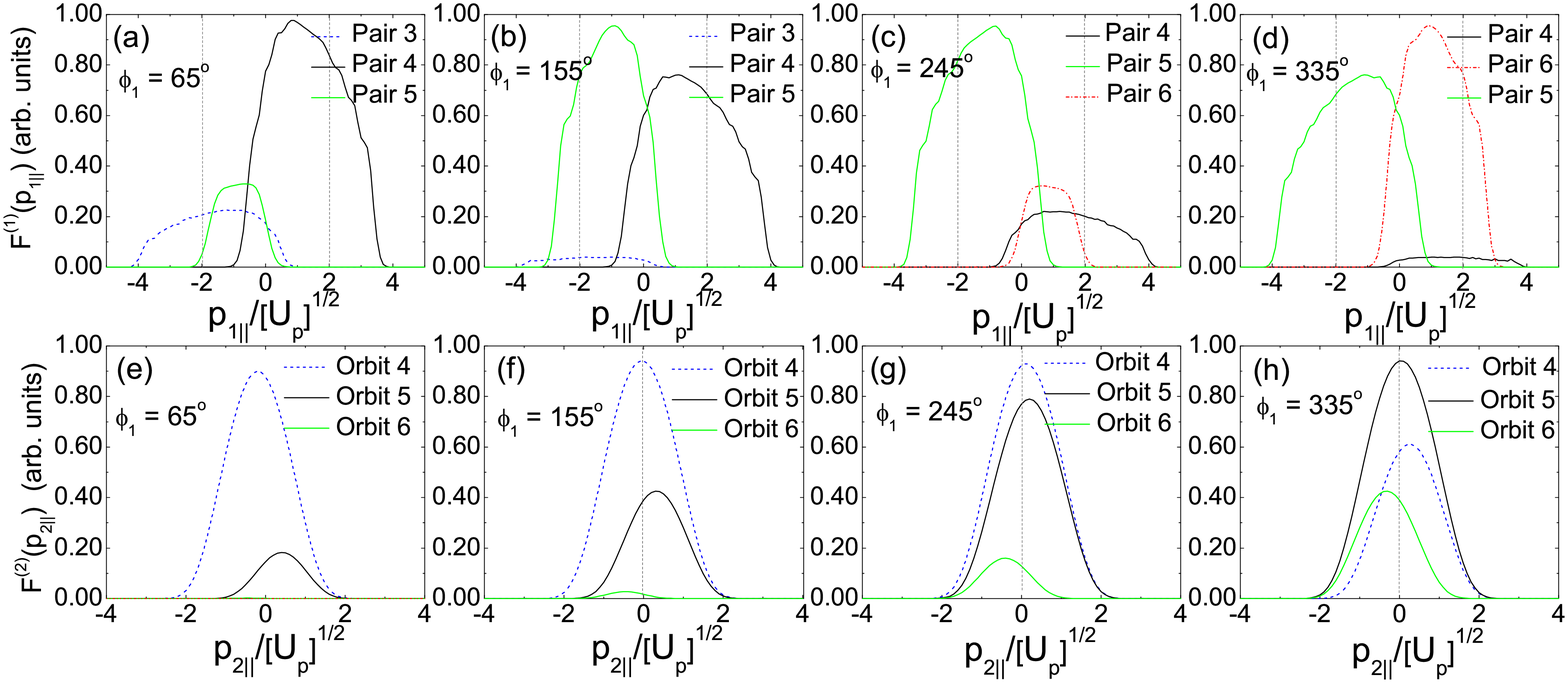} \caption{Partial electron momentum distributions $F^{(n)}(p_{n\parallel})$ $(n=1,2)$ (Eqs.~\ref{Mp1par} and \ref{Mp2par}) computed using the three most relevant individual ionization pathways for the first and second electron according to Figs.~\ref{Fig2} and \ref{Fig3} and Table \ref{relevantorbits}, for the same field and atomic parameters as in the previous figure. The upper and lower panels refer to the first and second electron, respectively. From left to right, the phases $\phi_1=65^{\circ}$ [panels (a) and (d)], $\phi_1=155^{\circ}$ [panels (b) and (e)], $\phi_1=245^{\circ}$ [panels (c) and (f)] and $\phi_1=335^{\circ}$ [panels (d) and (g)] have been taken. The probability densities have been normalized to slightly below unity for the dominant contributions in each panel, in order to facilitate a more direct comparison. The dotted lines in panels (a) to (d) indicate the momenta $\pm 2 \sqrt{U_p}$, where the partial distributions related to a monochromatic driving field are expected to be peaked.} \label{Partial}
\end{figure}

In Fig.~\ref{Partial}, we display $F^{(1)}(p_{1\parallel})$ and $F^{(2)}(p_{2\parallel})$ computed for \textit{each} of the contributions described in the previous section. For the first electron, the figure illustrates very clearly the loss of relevance in Pair $3(e_1)$ discussed above as the CEP increases (see Figs.~\ref{Partial}(a) and (b)). The contributions from such a pair, mostly located in the negative parallel momentum region, become vanishingly small already for $\phi_1=245^{\circ}$. In fact, even for $\phi_1=65^{\circ}$ this pair is not dominant, as the partial probability density associated with Pair $4(e_1)$, mostly in the positive momentum region, is higher.  This is a consequence of the fact that the field strength at the corresponding ionization time is higher for Pair $4(e_1)$, so that the first electron tunnels through a broader barrier (see Fig.~\ref{Fig2}). In Figs.~\ref{Partial}(b) to (d), the partial momentum distributions also show that Pair $5(e_1)$ is very important over a large CEP range, and in fact provides the dominant ionization pathway for the first electron for $\phi_1=155^{\circ}$ and $\phi_1=245^{\circ}$ (see Figs.~\ref{Partial}(b) and (c)). This is in agreement with with Fig.~\ref{Fig2}, which shows that, for a large range of phases, the instantaneous field at ionization is near its absolute maximum for this pair. In fact, only for $\phi_1=345^{\circ}=335^{\circ}$ does Pair $6(e_1)$ become dominant.
 Another interesting feature is that the curves corresponding to a specific Pair $n (e_1)$ are the mirror images of those related to Pair $ n+1 (e_1)$ if the CEP is shifted in $\Delta\phi=180^{\circ}$. This holds for all partial probabilities displayed in Figs.~\ref{Partial}(a) to (d). For instance, $F^{(1)}(p_{1\parallel})$ for Pair $3(e_1)$ in Fig.~\ref{Partial}(a) is equal to $F^{(1)}(-p_{1\parallel})$ for Pair $4(e_1)$ in Fig.~\ref{Partial}(c) and so on. Finally, the peaks of the partial distributions agree with the predictions in Sec.~\ref{orbitspulse} obtained from the solutions of the saddle-point equations.

 For the second electron, the partial probabilities $F^{(2)}(p_{2\parallel})$ confirm that the main ionization channel is via Orbit $4(e_1)$ for a large range of CEPs, as shown in Figs.~\ref{Partial}(e) and (f). The contributions from Orbit $5(e_2)$  only start to compete with this channel at $\phi_1=245^{\circ}$, and eventually become dominant at $\phi_1=335^{\circ}$ [Figs.~\ref{Partial}(g) and (h), respectively]. One should note that the symmetry  $F^{(2)}(p_{2\parallel})$ for Orbit $n(e_2)$  corresponds to $F^{(2)}(-p_{2\parallel})$ for Orbit $n+1 (e_2)$ also occurs upon a phase shift $\Delta\phi_1=180^{\circ}$. In the figure, however, it cannot be observed as Orbit $3(e_2)$ is missing. This orbit corresponds to ionization events triggered by Pairs $1(e_1)$ or $2(e_1)$, as illustrated in Figs.~\ref{Fig2} and \ref{Fig3}. Such pairs are irrelevant due to the weak field amplitudes involved\footnote{The probability density associated with Pair 2$(e_1)$ for $\phi_1=65^{\circ}$ is the mirror image of that related to Pair 3$(e_1)$ for $\phi=245^{\circ}$. We have verified that both rescattering events lead to a vanishingly small probability density.}.
In Table \ref{relevantorbits} we provide a summary of the most relevant orbits encountered for the first and second electron, in decreasing order of importance.
\begin{table}
    \begin{tabular}{ l|l l}
       \hline\hline
        $\phi_1$ (degrees) & First electron                                                           & Second electron                                                             \\ \hline
        $65^{\circ}$      &
  \begin{minipage}{3cm}
    \vskip 4pt
    \begin{enumerate}
   \item Pair 4 $(+)$
   \item Pair 3 $(-)$\\Pair 5 $(-)$
   \end{enumerate}
   \vskip 4pt
 \end{minipage}&
 \begin{minipage}{3cm}
    \vskip 4pt
    \begin{enumerate}
   \item Orbit 4
   \item Orbit 5
   \item Orbit 6
   \end{enumerate}
   \vskip 4pt
   \end{minipage}
   \\ \hline
     $155^{\circ}$      & \begin{minipage}{3cm}
    \vskip 4pt
    \begin{enumerate}
   \item Pair 5 $(-)$
   \item Pair 4 $(+)$
   \item Pair 3 $(-)$
   \end{enumerate}
   \vskip 4pt
 \end{minipage} &\begin{minipage}{3cm}
    \vskip 4pt
    \begin{enumerate}
   \item Orbit 4
   \item Orbit 5
   \item Orbit 6
   \end{enumerate}
   \vskip 4pt
   \end{minipage} %
       \\   \hline      $245^{\circ}$ &  \begin{minipage}{3cm}
    \vskip 4pt
    \begin{enumerate}
   \item Pair 5 $(-)$
   \item Pair 4 $(+)$\\ Pair 6 $(+)$
   \end{enumerate}
   \vskip 4pt
 \end{minipage}& \begin{minipage}{3cm}
    \vskip 4pt
    \begin{enumerate}
   \item Orbit 4\\
         Orbit 5
   \item Orbit 6
   \end{enumerate}
   \vskip 4pt
   \end{minipage}%
     \\  \hline $335^{\circ}$      &\begin{minipage}{3cm}
    \vskip 4pt
    \begin{enumerate}
   \item Pair 6 $(+)$
   \item Pair 5 $(-)$
   \item Pair 4 $(+)$
   \end{enumerate}
   \vskip 4pt
 \end{minipage}  &\begin{minipage}{3cm}
    \vskip 4pt
    \begin{enumerate}
   \item Orbit 5
   \item Orbit 4
   \item Orbit 6
   \end{enumerate}
   \vskip 4pt
   \end{minipage}\\
       \hline\hline
    \end{tabular}
    \caption{Most relevant orbits for the first and second electrons for the values of the carrier-envelope phase employed in this work, in order of decreasing importance. A single number for more than one orbit indicates that their contributions are comparable or competing, while different numbers indicate the clear dominance of a pair for the first electron, or an orbit for the second electron. The signs $(\pm)$ indicate whether a specific pair of orbits leads to electron momentum distributions $|F^{(1)}(p_{1\parallel})|^2$ peaked at positive $(+)$ or negative $(-)$ momentum. Subsequently, this will imply that the contributions triggered by such a pair of orbits will populate either he positive or the negative half axis in the parallel-momentum plane.}\label{relevantorbits}
\end{table}

\subsection{Correlated electron momentum distributions}
In view of the discussion presented above, one could conclude that a
pair of orbits will lead to dominant contributions in the
electron-momentum distributions if (i) the corresponding potential
barrier through which the electrons tunnel is as narrow as
possible, as this will lead to a high ionization probability; (ii) the
kinetic energy of the first electron upon return is as high as
possible, as this will populate a large region in momentum space.
One should bear in mind, however, that RESI is a correlated
two-electron process. This implies that the shapes and regions populated by the NSDI electron momentum distributions will be determined by the interplay between the dominant contributions from the first and second electron.

\begin{figure}[tbp]
\begin{center}
\noindent \includegraphics[width=10cm]{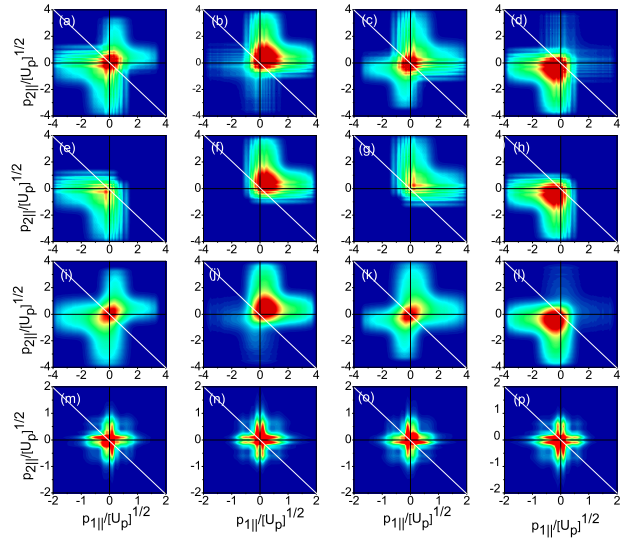}
\end{center}
\caption{Correlated RESI electron momentum distributions as functions of the parallel momentum components $p_{1\parallel}$, $p_{2\parallel}$ for the same atomic and field parameters as in the previous figures. In panels (a) to (l), constant ionization and excitation prefactors $V_{\mathbf{p}_1e, \mathbf{k}g}$, $V_{\mathbf{p}_2e}$ prefactors have been employed, while in panels (m) to (p) the prefactors related to the excitation from a $3p$ to a $4s$ state by a contact type interaction have been used. The explicit expressions for these prefactors is given in \cite{SNF2010}. From left to right, the carrier-envelope phases are $\phi_1=65^{\circ}$ [panels (a), (e), (i) and (m)], $\phi_1=155^{\circ}$ [panels (b), (f), (j) and (n)], $\phi_1=245^{\circ}$ [panels (c), (g), (k) and (o)], $\phi_1=335^{\circ}$ [panels (d), (h), (l) and (p)]. When constant prefactors are used, from the top to the bottom of the figure, we plot the distributions obtained as follows. Panels (a) to (d): all pairs of orbits for the first electron, and all possible ionization pathways for the second electron displayed in Table \ref{relevantorbits}. Panels (e) to (h): all ionization channels for the second electron specified in Table \ref{relevantorbits} and Pair $3(e_1)$ [panel (e)], $4 (e1)$ [panels (f) and (g)] and $5(e_1)$ [panel (h)] for the first electron. Panels (i) to (l): all pairs of orbits for the first electron specified in Table \ref{relevantorbits}, but only the ionization pathways for the second electron immediately after that specific ionization event, i.e., Orbit $4(e_2)$ for Pair $3(e_1)$, Orbit $5(e_2)$ for Pair $4(e_1)$ and Orbit $6(e_1)$ for Pair $5(e_1)$. In panels (m) to (o), we have included all orbits for the first and second electron specified in Table \ref{relevantorbits}. For panels (a) to (l), the color scales range from 0 to $1 \times 10^{-5}$, while in panels (m) to (o) they range from 0 to $1 \times 10 ^{-20}$. The white lines in the figure indicate the antidiagonal $p_{1\parallel}=-p_{2\parallel}$. }
\label{Fig7}
\end{figure}

In order to understand this issue, we compute the correlated electron-momentum distributions
\begin{equation}
F(p_{1\parallel },p_{2\parallel })=\int \hspace*{-0.1cm}\int \hspace*{0.0cm}%
d^{2}p_{1\perp }d^{2}p_{2\perp }|M(\mathbf{p}_{1},\mathbf{p}_{2})+\mathbf{p}%
_{1}\leftrightarrow \mathbf{p}_{2}|^{2},\label{distrp1p2}
\end{equation}%
as functions of the momentum components $p_{n\parallel}$ $(n=1,2)$ parallel to the laser-field polarization. In Eq.~(\ref{distrp1p2}), unless otherwise stated, $M(\mathbf{p}_{1},\mathbf{p}_{2})$ is the transition amplitude (\ref{Mp}) associated with the coherent sum of all processes over all sets of orbits, both for the first and the second electron. These transition amplitudes are symmetrized to account for the fact that both electrons are indistinguishable. The transverse-momentum components are integrated over. These distributions are plotted in Fig.~\ref{Fig7}.

For $\phi_1=65^{\circ}$, the distributions occupies a broad region along
the negative half axes $p_{i\parallel}=0$ $(p_{j\parallel}<0)$, with
$i\neq j$, as displayed in Fig.~\ref{Fig7}(a). This illustrates the importance of Pair $3(e_1)$,
as a direct comparison with Fig.~\ref{Fig7}(e) shows.  Even if this pair is not related to the most prominent ionization event for the second electron, upon rescattering, it triggers the dominant tunnel ionization channel for the second electron, i.e., along Orbit $4(e_2)$. This counterbalances the influence of Pair $4(e_1)$, which, according to the partial distributions in Fig. \ref{Partial} and our previous line of argument, is the prevalent ionization pathway for the first electron for this specific CEP value.  Contributions from Pair $4(e_1)$, along the parallel momenta
positive half axes are also present. These
contributions are, however, comparable, or even slightly weaker, as rescattering along Pair $4(e_1)$ can only lead to ionization along Orbit $5(e_2)$, Orbit $6 (e_2)$ or even later orbits.  Furthermore, because an electron returning along Pair $3(e_1)$ acquires a higher kinetic energy than if it returns along Pair $4(e_1)$, the momentum region populated by events related to former pair is larger. This fact is also observed for the partial momentum distribution in Fig.~\ref{Partial}(a), which exhibits comparable probability densities over a broad momentum region.

For $\phi=155^{\circ}$, in contrast, Pair $4(e_1)$ prevails (see Fig.~\ref{Fig7}(b) as compared to
\ref{Fig7}(f)). This leads to a shift in the distribution towards the positive half axis of the $p_{1\parallel}p_{2\parallel}$ plane.  Physically, this can be
attributed to the loss of relevance related to Pair $3(e_1)$, together with the fact that the ionization channel along Orbit $5(e_2)$ becomes more prominent. Hence, ionization along Orbit $4(e_2)$ can no longer counterweight the other effects.  This is in agreement with the previous discussions in Sec.~\ref{orbitspulse} and Fig.~\ref{Partial}. Note, however, that the partial probability density $F^{(1)}(p_{1\parallel})$ associated with Pair $5(e_1)$ is dominant in Fig.~\ref{Partial}(b). Nevertheless, Pair $5(e_1)$ can only lead to ionization events related to Orbit $6(e_2)$ or later. These events are too close to the pulse turn off to play a significant role.

The distributions obtained for $\phi_1=245^{\circ}$ [Fig.~\ref{Fig7}(c)], on the other hand, show that, as the CEP increases, Pair $5(e_1)$ becomes more relevant and starts to influence the overall distributions. For this specific phase, the ionization channel for first electron along Pair $3(e_1)$ is negligible, regardless of the subsequent ionization events. Ionization along Pair $4(e_1)$ is relatively small. It may, however, cause ionization of the second electron along Orbit $5(e_1)$, which is quite prominent. Pair 5$(e_1)$ is the most prominent ionization channel for the first electron, but may lead to ionization only along Orbit $6(e_2)$ or at later times. Hence, Pair $4(e_1)$, whose contributions are presented in Fig.~\ref{Fig7}(g), still determines the momentum regions to be occupied. Interestingly, this distribution is the mirror image of that obtained for $\phi_1=65^{\circ}$ with regard to $(p_{1\parallel},p_{2\parallel})\rightarrow (-p_{1\parallel},-p_{2\parallel})$.

Finally, for $\phi_1=335^{\circ}$, displayed in Fig.~\ref{Fig7}(d), the distributions are almost entirely concentrated along the negative half axis  $p_{i\parallel}=0$ $(p_{j\parallel}<0)$, with $i\neq j$. This is a consequence of the fact that, together, the pathways related to Pair $5(e_1)$ for the first electron, and Orbit $6(e_2)$ for the second electron determine the momentum regions to be populated (see Fig.~\ref{Fig7}(h) for comparison). A similar interpretation provided when discussing the distributions obtained in Fig.~\ref{Fig7}(a), for $\phi_1=155^{\circ}$, applies, with the difference that all the indices must be shifted by 2, i.e., instead of Pairs $3(e_1)$ and $4(e_1)$, now one must consider Pairs $5(e_1)$ and $6(e_1)$ and the subsequent orbits for the second electron. Note once more that this distribution is the mirror image of that depicted in Fig.~\ref{Fig7}(b).

Apart from the effects discussed above, throughout, there exist residual fringes parallel to the axis $p_{n\parallel}=0$. These fringes are related to the fact that, in our computations, we have included all the relevant orbits along which the second electron may tunnel, subsequently to being excited by a particular pair of orbits. Quantum mechanically, the transition amplitudes related to these pathways interfere and even partly survive the integration over the transverse momentum components. In a more realistic scenario, however, we expect these fringes to be absent, due to the fact that the excited bound state from which the second electron is released is strongly depleted. Hence, ionization would mainly occur around the field maximum closest to the time of excitation.

In order to mimic depletion, in Figs.~\ref{Fig7}(i) to (l) we consider only this ionization pathway. As an overall feature, the above-mentioned structure disappears. Furthermore, the correlated probability densities are now slightly displaced from the axes $p_{n\parallel}=0$. This occurs because, due the lack of monochromaticity of the few-cycle pulse, $\mathrm{Im}[\omega t]$ no longer exhibits a minimum at $p_{\parallel}=0$ (see discussion in Sec.~\ref{orbitspulse}). If many ionization events are considered, this effect tends to average out and this asymmetry is less prominent.  Concretely, the upper parts of the distributions in Fig.~\ref{Fig7} are slightly displaced towards the first quadrant of the parallel momentum place. This is caused by the fact that ionization along Orbit $5(e_2)$, which is the main ionization channel for the second electron related to Pairs $4(e_1)$ and $6(e_1)$, is favored for positive momenta (see Fig.~\ref{Fig5}(f)) and subsequent discussion). In contrast, the lower parts of such distributions are slightly shifted towards the third quadrant. This is due to the fact that the main ionization channel for the second electron in this momentum region is either along Orbit $4(e_2)$ ($\phi_1=65^{\circ}$ and $\phi_1=155^{\circ}$) or along Orbit $6(e_2)$ ($\phi_1=245^{\circ}$ and $\phi_1=335^{\circ}$). For both orbits, ionization is slightly enhanced for negative momenta, as shown in Figs.~\ref{Fig5}(e) and (g).
Finally, in Figs.~\ref{Fig7}(m) to (p) we incorporate the prefactors $V_{\mathbf{p}_1g,\mathbf{k}g}$ and $V_{\mathbf{p}_2e}$ corresponding to an excitation process from the $3p$ to the $4s$ state  a contact-type interaction $V_{12}(\mathbf{r}_1,\mathbf{r}_2)=\delta(\mathbf{r}_1-\mathbf{r}_2)$. For the explicit expressions see Ref.~\cite{SNF2010}. In general, these prefactors introduce a bias towards low momenta, so that the distributions are much more focused around the origin $p_{1\parallel}=p_{2\parallel}=0$. For a Coulomb-type interaction $V_{12}=1/|\mathbf{r}_2-\mathbf{r}_1|$, we have verified that even lower momenta are favored.
\section{Conclusions}
\label{Conclusions}
In this work, we highlight the influence of the carrier-envelope phase (CEP) for the recollision-excitation with subsequent tunneling ionization (RESI) pathway in NSDI with few-cycle pulses. The electron-momentum distributions are quite sensitive with regard to the CEP, and, as this parameter varies, move from the region below to the region above the anti-diagonal $p_{1\parallel}=-p_{2\parallel}$, or vice versa, in the plane spanned by the electron-momentum components parallel to the laser field polarization. Similarly to our previous studies in \cite{NSDI2004_Faria,NSDI2004_2_Faria} performed for the simpler NSDI mechanism of electron-impact ionization, we show that all features encountered can be explained in terms of electron trajectories. Indeed, a detailed analysis of where these trajectories are located within the pulse, and, in the context of the steepest descent method, of the real and imaginary parts of the ionization times of both electrons and the rescattering time of the first electron provide a consistent picture related to the shapes, maxima and regions populated by the RESI electron-momentum distributions. All such features are determined by the interplay between the dominant sets of trajectories for the first and second electron.

Qualitatively, the results in the present publication agree with those observed experimentally in \cite{Bergues2011}, in which it has been reported that, depending on the CEP, the electron-momentum distributions shift from the region below to the region above the anti-diagonal $p_{1\parallel}=-p_{2\parallel}$.  Quantitatively, however, we find a much larger probability density around the origin $p_{1\parallel}=p_{2\parallel}=0$ than those encountered in \cite{Bergues2011}. This is particularly true if we consider the prefactors $V_{\mathbf{p}_2e}$, $V_{\mathbf{p}_1e,\mathbf{k}g}$ associated with an excitation from the $3p$ to the $4s$ state in Argon. In this latter case, the distributions become much more localized around the origin.
These discrepancies may be related to several issues. First, there is always an uncertainty in the experimentally measured peak intensities. If this intensity is smaller than that employed in our computations, this means that the region around $p_{1\parallel}=p_{2\parallel}=0$ is less populated than in Fig. \ref{Fig7} (see \cite{SF2010,SNF2010} for details). Second, in a more realistic scenario, it may be that other excitation processes than that considered in this work play an important role. This implies that the features specific to particular bound states may average out, and one may approach the results obtained when the prefactors were left out. Third, the momentum constraints related to the first and the second electron, which have been derived within the context of the strong-field approximation \cite{SNF2010,SFS2012} neglect the presence of the binding potential when the electron is in the continuum, and as well bound-state depletion. Both issues may affect such constraints.

Finally, we would like to comment on the role of depletion on the RESI electron-momentum distributions. In \cite{Bergues2011}, the probability densities at nonvanishing momenta which led to the axes of the cross have been attributed to the second electron. It has been argued that, due to depletion, the second electron left before the peak field with non-vanishing momentum. Our results, together with the constraints stated in this paper, suggest that depletion will mainly shift the probability density with regard to the axis $p_{n\parallel}=0$ (see discussion of panels (i) to (l) in Fig.~\ref{Fig7}). This effect, however, seems to be relatively small. According to our model, the shift in the probability density away from the origin $p_{1\parallel}=p_{2\parallel}$ is mainly caused by the first electron, which, upon recollision, acquires the additional momentum $-A(t)$ from the field according to the constraints stated in Sec.~\ref{theory}. The issue of depletion is not yet fully understood, and will be pursued in future work.

\textbf{Acknowledgements:} We are grateful to B. Bergues and M. Kling for showing us their experimental data prior to publication, and to the Max Planck Institute for Quantum Optics, Munich, for their kind hospitality. This work was partly funded by the EPSRC/UCL PhD+ scheme.

\end{document}